\documentclass{aa}
\usepackage{graphicx}
\usepackage{txfonts}
\usepackage{ulem}

\usepackage[colorlinks=true,citecolor=blue]{hyperref}
\usepackage{natbib}
\begin{document} 

   \title{The GTC exoplanet transit spectroscopy survey. VI.
         \thanks{Based on observations made with the Gran Telescopio Canarias (GTC), 
         at the Spanish Observatorio del Roque de los Muchachos of the Instituto de 
         Astrof\'{i}sica de Canarias, on the island of La Palma, as well as observations
         obtained at the European Southern Observatory at Paranal, Chile in
         program 096.C-0258(A)}}

   \subtitle{A spectrally-resolved Rayleigh scattering slope in GJ 3470b}

   \author{G. Chen\inst{1,2,3}
          \and
          E.~W. Guenther\inst{4,5}
          \and
          E. Pall\'{e}\inst{1,2}
          \and
          L. Nortmann\inst{1,2}
          \and
          G. Nowak\inst{1,2}
          \and
          S. Kunz\inst{4}
          \and
          H. Parviainen\inst{1,2}
          \and
          F. Murgas\inst{1,2}
          }

   \institute{Instituto de Astrof\'{i}sica de Canarias, V\'{i}a L\'{a}ctea s/n, E-38205 La Laguna, Tenerife, Spain\\
         \email{gchen@iac.es}
         \and
             Departamento de Astrof\'{i}sica, Universidad de La Laguna, Spain
         \and
             Key Laboratory of Planetary Sciences, Purple Mountain Observatory, Chinese Academy of Sciences, Nanjing 210008, China
         \and
             Th\"uringer Landessternwarte Tautenburg, Sternwarte 5, 07778, Tautenburg, Germany
         \and
             Visiting scientist Instituto de Astrof\'{i}sica de Canarias, V\'{i}a L\'{a}ctea s/n, E-38205 La Laguna, Tenerife, Spain
             }

   \date{Received December 12, 2016; accepted March 6, 2017}

 
  \abstract
{}
{As a sub-Uranus-mass low-density planet, GJ 3470b has been found to show a flat featureless transmission spectrum in the infrared and a tentative Rayleigh scattering slope in the optical. We conducted an optical transmission spectroscopy project to assess the impacts of stellar activity and to determine whether or not GJ 3470b hosts a hydrogen-rich gas envelop.}
{We observed three transits with the low-resolution OSIRIS spectrograph at the 10.4~m Gran Telescopio Canarias, and one transit with the high-resolution UVES spectrograph at the 8.2~m Very Large Telescope. }
{From the high-resolution data, we find that the difference of the \ion{Ca}{II}\,H+K lines in- and out-of-transit is only $0.67\pm0.22\%$, and determine a magnetic filling factor of about 10--15\%. From the low-resolution data, we present the first optical transmission spectrum in the 435--755~nm band, which shows a slope consistent with Rayleigh scattering.}
{After exploring the potential impacts of stellar activity in our observations, we confirm that Rayleigh scattering in an extended hydrogen/helium atmosphere is currently the best explanation. Further high-precision observations that simultaneously cover optical and infrared bands are required to answer whether or not clouds and hazes exist at high-altitude.}

   \keywords{Planetary systems --
             Planets and satellites: individual: GJ 3470b --
             Planets and satellites: atmospheres --
             Techniques: spectroscopic}

   \maketitle

\section{Introduction}
\label{sectI}

Transiting planets have become an invaluable population for atmospheric characterization since the first discovery \citep{2000ApJ...529L..45C,2000ApJ...529L..41H}. The unique transit geometry enables the observations of transmission spectroscopy \citep[e.g.][]{2002ApJ...568..377C}, thermal emission \citep[e.g.][]{2005Natur.434..740D,2005ApJ...626..523C}, or phase curve \citep[e.g.][]{2007Natur.447..183K} originating from planetary atmospheres. Given that thicker hydrogen/helium (H/He) gas envelope and higher temperature could potentially produce larger atmospheric spectral signatures, a large number of hot Jupiters have been observed by the {\it Hubble space telescope} ({\it HST}) and {\it Spitzer space telescope} for their atmospheres \citep[e.g.][]{2010ARA&A..48..631S,2014PASA...31...43B}. The well-studied hot-Jupiter sample has not only resulted in robust detection of the sodium (Na) and potassium (K) atoms and the water (H$_2$O) molecule, but also revealed a continuum from clear to cloudy atmospheres \citep{2016Natur.529...59S}.

NASA's {\it Kepler} mission has shown that small planets (e.g. $\lesssim$4$R_\oplus$) are the most common planets around Sun-like stars \citep{2011ApJ...742...38Y,2012ApJS..201...15H,2013ApJ...766...81F} as well as M-type stars \citep{2013ApJ...767...95D}. The atmospheres of small planets are more complex than hot Jupiters given the transition from accreted thick gas envelopes to outgassed thin layers. Most of the current observations show the commonness of flat and featureless near-infrared transmission spectrum in small planets, including \object{GJ 1214b} \citep{2010Natur.468..669B,2011ApJ...743...92B,2012ApJ...747...35B}, \object{GJ 436b} \citep{2014Natur.505...66K}, \object{HD 97658b} \citep{2014ApJ...794..155K}, \object{GJ 3470b} \citep{2014A&A...570A..89E}, and \object{TRAPPIST-1b+c} \citep{2016Natur.537...69D}, indicative of either high-altitude thick clouds or high atmospheric mean molecular weight. Based on extensive repeated transit observations on the super-Earth GJ 1214b, \citet{2014Natur.505...69K} reported the conclusive inference of clouds. On the other hand, the Neptune-sized planet \object{HAT-P-11b} was found to show water vapor absorption at the wavelength 1.4~$\mu$m \citep{2014Natur.513..526F}, while the super-Earth \object{55 Cnc~e} might have a H-rich atmosphere with HCN \citep{2016ApJ...820...99T}.

The sub-Uranus-mass planet \object{GJ 3470b} was discovered to transit an M1.5V star every 3.34~days at an orbital distance of 0.03~AU by \citet{2012A&A...546A..27B}. Its low bulk density \citep[$13.73\pm 1.61 M_\oplus$, $3.88\pm 0.32 R_\oplus$;][]{2014MNRAS.443.1810B} indicates that significant amount of hydrogen and helium gases should be present \citep{2013ApJ...768..154D}, which cannot be formed by outgassing alone according to \citet{2011ApJ...738...59R}. This planet also has a low surface gravity $\log g_{\rm p}=2.83\pm 0.11$ and a relatively warm equilibrium temperature $T_{\rm eq}=(1-A_{\rm B})^{1/4}(692\pm 15)$~K \citep{2014MNRAS.443.1810B}, where $A_{\rm B}$ is the Bond albedo. Given its proximity to the Earth, the host star is very bright \citep[$g'{\rm mag}=13.0$, $r'{\rm mag}=11.7$, $i'{\rm mag}=10.7$;][]{2015AJ....150..101Z}. These favorable conditions make \object{GJ 3470b} one of the most important targets for the atmospheric characterization of small planets. 

\citet{2013ApJ...770...95F} performed simultaneous observations in the $g'$, $R_{\rm c}$, $I_{\rm c}$, and $J$ bands with the 50~cm and 188~cm telescopes at the Okayama Astrophysical Observatory, and found that the planet radius is $5.8\pm 2.0$\% larger in the $I_{\rm c}$ band than in the $J$ band. \citet{2013A&A...559A..32N} analyzed their high-quality light curves simultaneously obtained by two LBC cameras at the Large Binocular Telescope (LBT), and found that the planet radius is $9.7\pm 1.9$\% larger in the ultraviolet ($\lambda_{\rm c}=357.5$~nm) than in the red-optical ($\lambda_{\rm c}=963.5$~nm), which together with \citet{2013ApJ...770...95F}'s measurements were interpreted as a signature of scattering processes. Since then, several follow-up multi-epoch optical photometric studies \citep{2014MNRAS.443.1810B,2015ApJ...814..102D,2016MNRAS.463.2574A} have reported tentative evidence of Rayleigh scattering in \object{GJ 3470b}'s atmosphere. 

\citet{2013A&A...559A..33C} performed the first transmission spectroscopy for \object{GJ 3470b} in the 2.09--2.36~$\mu$m band with the MOSFIRE spectrograph at the Keck telescope, and obtained a flat transmission spectrum. Using the WFC3 instrument on {\it HST}, \citet{2014A&A...570A..89E} measured a flat transmission spectrum in the 1.1--1.7~$\mu$m band, which also agrees with the {\it Spitzer} measurement at 4.5~$\mu$m \citep{2013ApJ...768..154D}. Together with the optical measurements in the literature, they ruled out H-rich atmospheres that are cloud-free or with tholin hazes, and suggested a cloudy H-rich atmosphere to explain this dichromatic transmission spectrum from 0.3 to 5.0~$\mu$m. 

However, most of previous optical measurements were low-quality broad-band photometry obtained at different epochs. It is still possible that the tentatively observed slope that mimics Rayleigh scattering actually comes from the contamination of stellar activity \citep[e.g.][]{2014A&A...568A..99O}. We therefore conducted a program that includes both high- and low-resolution optical spectroscopy to acquire \object{GJ 3470b}'s transmission spectrum. This enables us to assess the impacts of stellar activity, and to determine whether or not \object{GJ 3470b} has a H-rich gas envelop. 

This paper is organized as follows. In Sections~\ref{sectII} and \ref{sectIII}, we will describe the observations, data reduction and the results obtained from the low-resolution GTC/OSIRIS observations and the high-resolution VLT/UVES observations, respectively. In Section~\ref{sectIV}, we will discuss the impacts of stellar activity. In Section~\ref{sectV}, we will interpret the observed planetary atmosphere. Finally, we will give our conclusions in Section~\ref{sectVI}.

\section{Low-resolution spectra}
\label{sectII}

\subsection{GTC/OSIRIS transit observations}

Using the Optical System for Imaging and low-Intermediate-Resolution Integrated Spectroscopy \citep[OSIRIS;][]{2012SPIE.8446E..4TS} at the Nasmyth-B focal station of the 10.4~m Gran Telescopio Canarias (GTC), we observed three transits of the warm sub-Uranus-mass planet \object{GJ 3470b} on the nights of January 30, February 9, and March 10 in 2016. For all three observations, OSIRIS was configured in the long-slit spectroscopic mode with the R1000B grism. In each exposure, \object{GJ 3470} ($r'{\rm mag}=11.7$) was observed simultaneously with the reference star 2MASS J07591316+1525479 ($r'{\rm mag}=10.4$; 2.9$'$ away) in the same slit. The resulting spectral images were recorded by two red-optimized $2048\times4096$ Marconi CCDs in the 200 kHz readout mode with the $2\times2$ binning, which provided an unvignetted field of view (FOV) of 7.4 arcmin in the spatial direction and a pixel scale of 0.254$''$. The overheads between two exposures are $\sim$23.5 seconds. A gap of 9.4$''$ exists between these two CCDs. With an instrumental dispersion of 2.1~\AA~~per pixel, the R1000B grism gives a wavelength range from 365 to 775 nm. The overall observing logs are summarized in Table~\ref{tab:osiris_obslog}.

\begin{table*}
  \centering
  \small
  \caption{Observing summary}
  \label{tab:osiris_obslog}
  \begin{tabular}{ccccccccc}
    \hline\hline\noalign{\smallskip}
    Run & Date & Slit   & Observing time & $T_{\rm exp}$ & $N_{\rm obs}$ & Airmass & Rotator angle & Seeing\\\noalign{\smallskip}
    (\#) &      & ($''$) & (UT)     & (s) &           &         &  ($^\circ$) & ($''$)\\\noalign{\smallskip}
    \hline\noalign{\smallskip}
    \multicolumn{9}{c}{$GTC/OSIRIS$}\dotfill\\\noalign{\smallskip}
    1 & 2016-01-30 & 12 & 21:01--01:31 & 20 & 347 & 1.57$\rightarrow$1.03$\rightarrow$1.06 & 23$\rightarrow$37$\rightarrow$$-$53 & 1.9--10.2\\\noalign{\smallskip}
    2 & 2016-02-09 & 40 & 21:23--01:00 & 7 & 425 & 1.25$\rightarrow$1.03$\rightarrow$1.07 & 33$\rightarrow$37$\rightarrow$$-$59 & 0.8--1.6\\\noalign{\smallskip}
    3 & 2016-03-10 & 40 & 22:01--01:35 & 7 & 421 & 1.03$\rightarrow$1.63 & $-$11$\rightarrow$$-$110 & 0.9--2.5\\\noalign{\smallskip}
    \hline\noalign{\smallskip}
    \multicolumn{9}{c}{$VLT/UVES$}\dotfill\\\noalign{\smallskip}
    4 & 2016-01-14 & 0.8 & 04:50--08:40 & 568 & 22 & 1.31$\rightarrow$2.22 & 179$\rightarrow$128 & 0.7--1.5\\\noalign{\smallskip}
    \hline
  \end{tabular}
\end{table*}

\subsubsection{Run 1: the transit on January 30, 2016}

The 12$''$-wide slit was employed. Both \object{GJ 3470} and its reference star were placed on the CCD chip 1. Their centroids were roughly 21$''$ and 10$''$ away from the slit edges in the spatial direction, respectively. Due to technical problems, the time within 23:07--23:23 UT was lost. The HeAr and Ne arc lamps were measured through the 1.23$''$ slit. The weather condition was bad. The seeing was highly variable, ranging from 1.9$''$ to 10.2$''$ with a median value of 4.4$''$. Due to low data quality and unknown impact from the highly variable and large seeings (compared to 12$''$ narrow slit), we decided not to include this data set in any analysis except for the mid-transit time.

\subsubsection{Run 2: the transit on February 9, 2016}

\begin{figure}
 \centering
 \includegraphics[width=0.9\linewidth,angle=0.0]{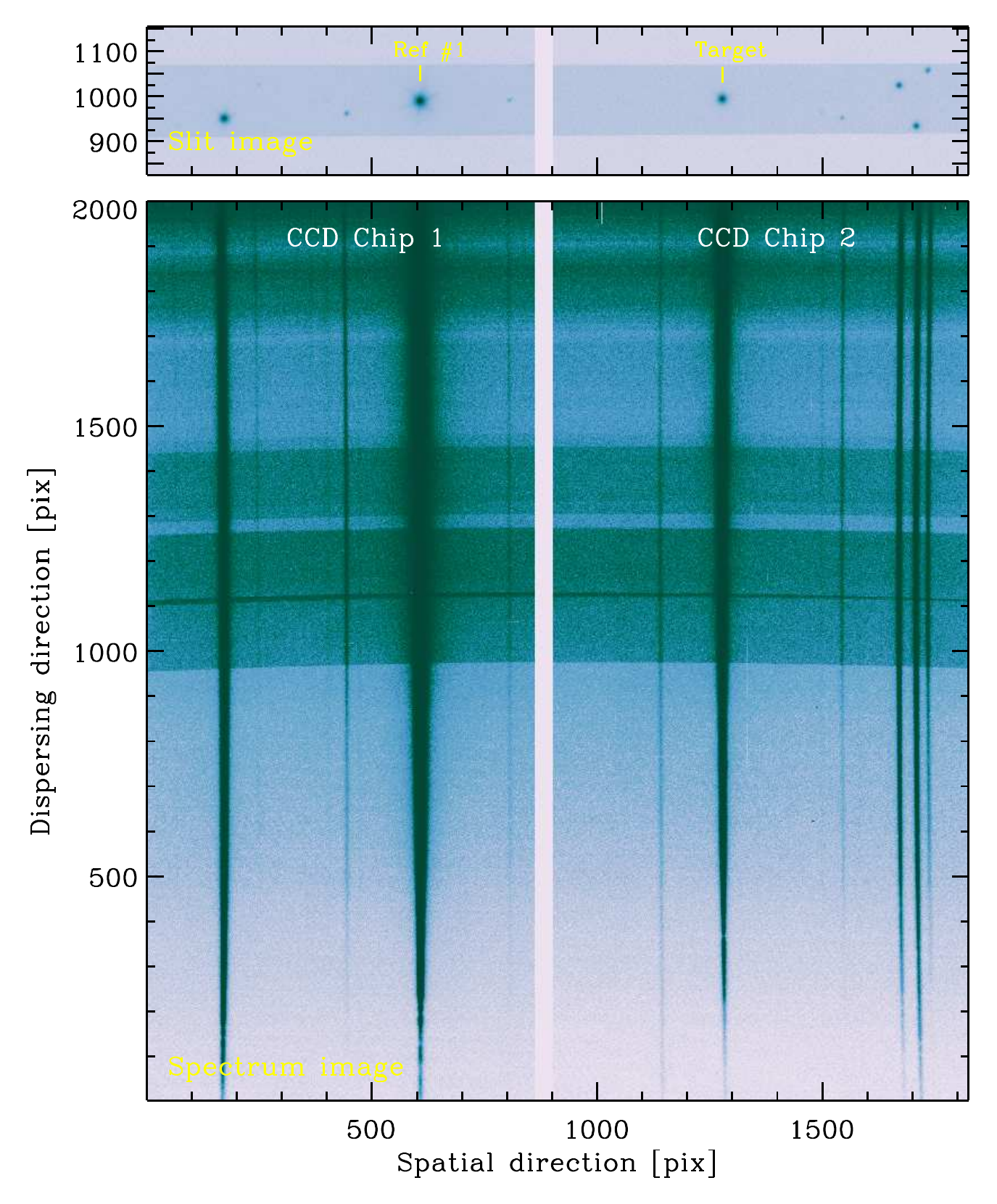}
 \caption{Acquisition images through the 40$''$ slit ({\it top panel}) and corresponding dispersed two-dimensional spectra ({\it bottom panel}) for Runs 2 and 3 obtained with GTC/OSIRIS. Note that Run 1 had a slightly different pointing, where both Ref \#1 and Target were placed on CCD chip 1 and the 12$''$ slit was used. \label{fig:gtc_slit_image}}
\end{figure}

The 40$''$-wide slit was employed. With the lessons learned in Run 1, \object{GJ 3470} and its reference star were placed on CCD chips 2 and 1, respectively, to avoid possible slit losses on the slit edges (see Fig.~\ref{fig:gtc_slit_image}). The night was clear. No moon was present during this observation. The seeing had a range of 0.8--1.6$''$, with a median value of 1.0$''$. The seeing-limited spectral resolution was $\sim$8~\AA. The position drifts caused by instrumental flexure and atmospheric refraction were determined by fitting Gaussian function to stellar spatial profiles and stellar absorption lines. The position drift in the spatial direction was measured to be less than 1 pixel, while it was not noticeable in the dispersing direction. The HeAr and Ne arc lamps were measured through the 1.0$''$ slit. 

\subsubsection{Run 3: the transit on March 10, 2016}

This Run had the same instrument configuration as Run 2. The 40$''$-wide slit was employed, and two stars were placed on different CCD chips. No moon was present during this observation. The seeing ranged from 0.9$''$ to 2.5$''$, with a median value of 1.5$''$. The seeing-limited spectral resolution was $\sim$12~\AA. The position drifts in the spatial and dispersing directions were measured to be within 1 pixel and 0.5 pixel, respectively. The HeAr and Ne arc lamps were measured through the 0.8$''$ slit. 

\subsection{Data reduction}

We reduced the acquired GTC/OSIRIS data using the approach described in \citet{chen2016a}, which makes use of the standard IRAF\footnote{IRAF is distributed by the National Optical Astronomy Observatory, which is operated by the Association of Universities for Research in Astronomy (AURA) under a cooperative agreement with the National Science Foundation.} routines and customized IDL\footnote{IDL stands for Interactive Data Language. It is a registered trademark of Exelis Inc. For further details see: \url{http://www.exelisvis.com/ProductsServices/IDL.aspx}.} scripts. The data reduction process included trimming of overscan region, subtraction of bias structure, correction of flat field, and removal of sky background. For each night, around 40 to 50 individual bias and flat measurements were used to create the master calibration files, respectively. A two-dimensional pixel-to-wavelength transformation map was constructed based on the line identification of the HeAr and Ne arc lamps. After the stars were masked out and the spectral image was mapped into the wavelength space, a sky background model was constructed for each exposure, and then mapped back into the pixel space, where the sky level under the masked star regions was interpolated wavelength-by-wavelength. The cosmic-ray hits were removed by a simple sigma-clipping method in the time domain for each pixel. 

The one-dimensional (1D) spectra were extracted using the optimal extraction algorithm \citep{1986PASP...98..609H}. The aperture size was fixed for all the exposures in a given Run. The optimal aperture was determined by minimizing the standard deviation of the white-color light-curve residuals among the data sets created with various aperture sizes. The adopted aperture diameters were 57 pixels for Run 1, 29 pixels for Run 2, and 22 pixels for Run 3. As the aperture size increases, the aperture-scatter growth curve always drops quickly at very small apertures, and then rises slowly after reaching the lowest scatter. We confirmed that the aperture choices did not affect our results when the aperture-scatter growth curve became stable. 

The wavelength solutions for the extracted 1D spectra was calculated by fitting a third order B-spline function\footnote{See \url{http://www.sdss.org/dr12/software/idlutils/}.} to the identified arc lines. For a given star, the spectrum of every exposure was cross-correlated with the first exposure to correct position drift in time. The wavelength difference between the target and reference spectra in the same exposure was corrected as well. Furthermore, the wavelength solutions were corrected to match the air wavelengths of specific stellar absorption lines in the still frame. We note that all the corrections were made in the wavelength solutions, instead of directly interpolating the spectra. 

\begin{figure}
 \centering
 \includegraphics[width=\linewidth,angle=0.0]{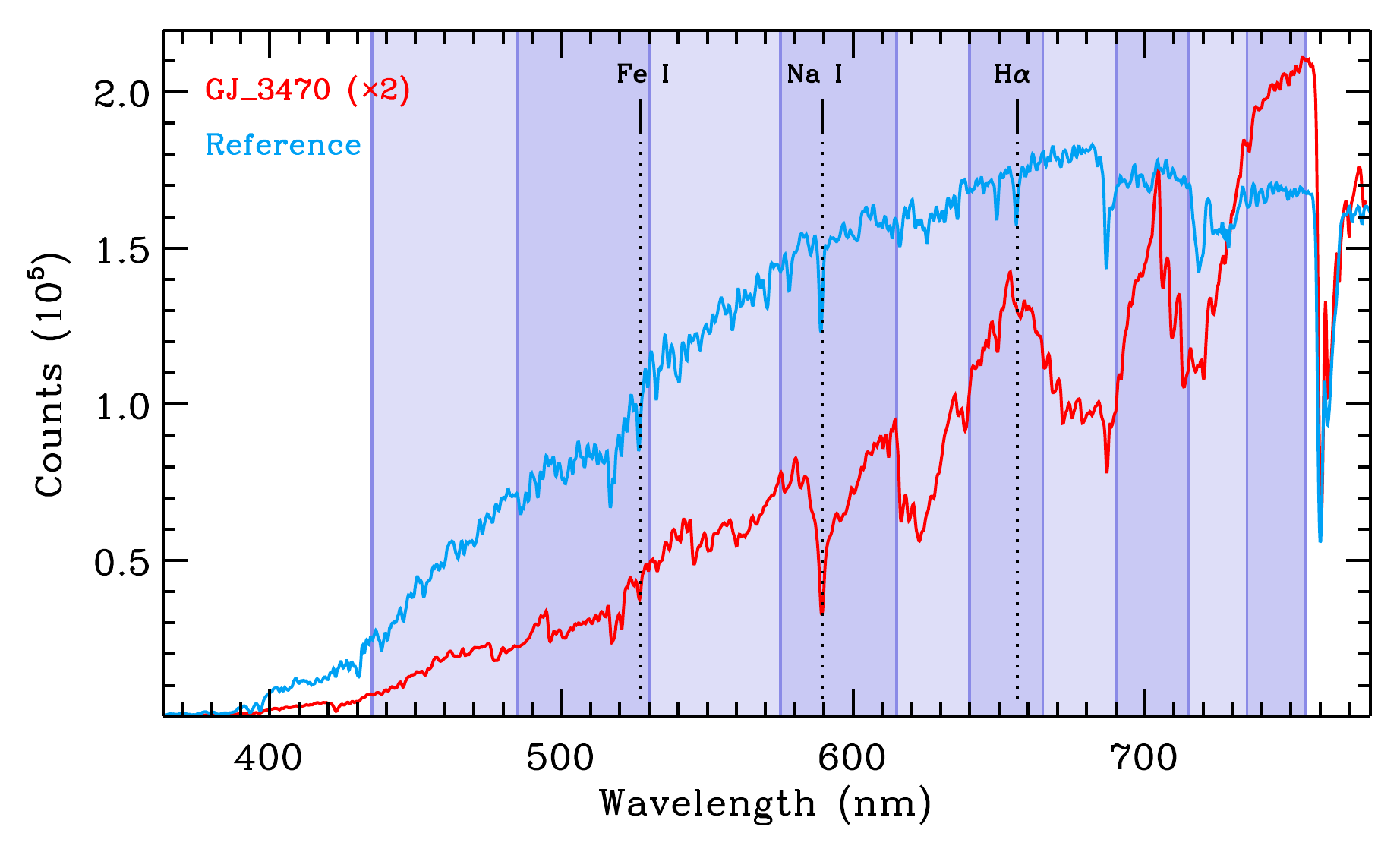}
 \caption{Example stellar spectra of \object{GJ 3470} (red) and its reference star (blue) obtained with GTC/OSIRIS. The color-shaded areas indicate the divided passbands that are used to create spectroscopic light curves. \label{fig:gtc_spec}}
\end{figure}

The time stamp was extracted from the FITS headers and shifted to the mid-point of each exposure. They were converted to the Barycentric Julian Date in the Barycentric Dynamical Time standard ($\mathrm{BJD}_\mathrm{TDB}$) using the IDL procedures written by \citet{2010PASP..122..935E}. To create a light curve, the pixel range of a requested passband was calculated using the corrected wavelength solution for each exposure. The counts of the complete pixels within this range were directly summed, while those of the edge pixels were fractionally added. After dividing the target flux by the reference flux, the resulting flux ratios were normalized by the out-of-transit level to create the final light-curve products. Finally, we created the white-color light curve in the wavelength range from 435~nm to 755~nm, and a set of spectroscopic light curves whose passbands are shown as shaded area in Fig.~\ref{fig:gtc_spec}. Since the flux level significantly dropped when approaching the shorter wavelength, the spectroscopic passbands were not evenly divided on purpose, to increase the signal-to-noise ratios in the bluer channels.  

\subsection{Light-curve analysis}
\label{sec:general_analysis}

We fit the GTC/OSIRIS transit light curves in the same procedures detailed in \citet{chen2016a}. The only difference here is that no third-light contamination exists. Therefore, dilution correction is removed from the fitting procedure. The light-curve data were directly fit with the \citet{2002ApJ...580L.171M} analytic transit model $\mathcal{T}(p_i)$ multiplied by a systematic decorrelation baseline model $\mathcal{B}(c_j)$. The potential free parameters in the transit model $\mathcal{T}(p_i)$ included the mid-transit time $T_{\rm mid}$, the orbital inclination $i$, the scaled semi-major axis $a/R_\star$, the planet-to-star radius ratio $R_{\rm p}/R_\star$, and the quadratic limb-darkening coefficients ($u_1$, $u_2$). The orbital period $P$ was always fixed to the literature value or the revised value as given below. Since \citet{2012A&A...546A..27B} placed a 1-$\sigma$ upper limit of 0.051 on the orbital eccentricity, the circular orbit was adopted in our analysis. On the other hand, the potential free parameters in the baseline model $\mathcal{B}(c_j)$ were the coefficients $c_j$ for the polynomial combinations of the state vectors $\vec{s}$. 

The limb-darkening coefficients $u_1$ and $u_2$ were fitted with Gaussian priors. Their theoretical values were interpolated from the ATLAS stellar atmosphere models using the stellar effective temperature $T_{\rm eff}=3652\pm 50$~K, surface gravity $\log g=4.78\pm 0.12$, and metallicity ${\rm [Fe/H]}=0.17\pm 0.06$ \citep{2014MNRAS.443.1810B}. The interpolation was performed using the Python package written by \citet{2015MNRAS.450.1879E}, where the stellar intensity profiles had been interpolated onto evenly spaced 100-$\mu$ ($\mu=\cos\theta$) grids \citep{2011A&A...529A..75C}. The bandpass response and telluric absorption features have been taken into account in the interpolation. The imposed prior widths were $\sigma_u=0.1$, which can make use of the known physical information of the host star without imposing too strict constraints.

The initial best-fitting model parameters were determined by the \texttt{MPFIT} package \citep{2009ASPC..411..251M}. A customized version of the Transit Analysis Package \citep[TAP;][]{2012AdAst2012E..30G} was employed to perform the Markov Chain Monte Carlo (MCMC) analysis to explore the likelihood distributions of the fitted parameters. This IDL package implements a Metropolis-Hastings MCMC technique within a Gibbs sampler \citep{2005AJ....129.1706F,2006ApJ...642..505F}. To account for the correlated noise, TAP uses a likelihood function \citep{2009ApJ...704...51C} that introduces the Daubechies fourth-order wavelet decomposition and assumes the power spectral density varies as $1/f$ at the frequency $f$. The final reported parameter values and uncertainties were calculated as the median and 1-$\sigma$ percentiles of each parameter's likelihood distribution. 

\subsubsection{Fitting of white-color light curves}
\label{sec:fit_white}

\begin{figure*}
 \centering
 \includegraphics[width=1.\linewidth,angle=0.0]{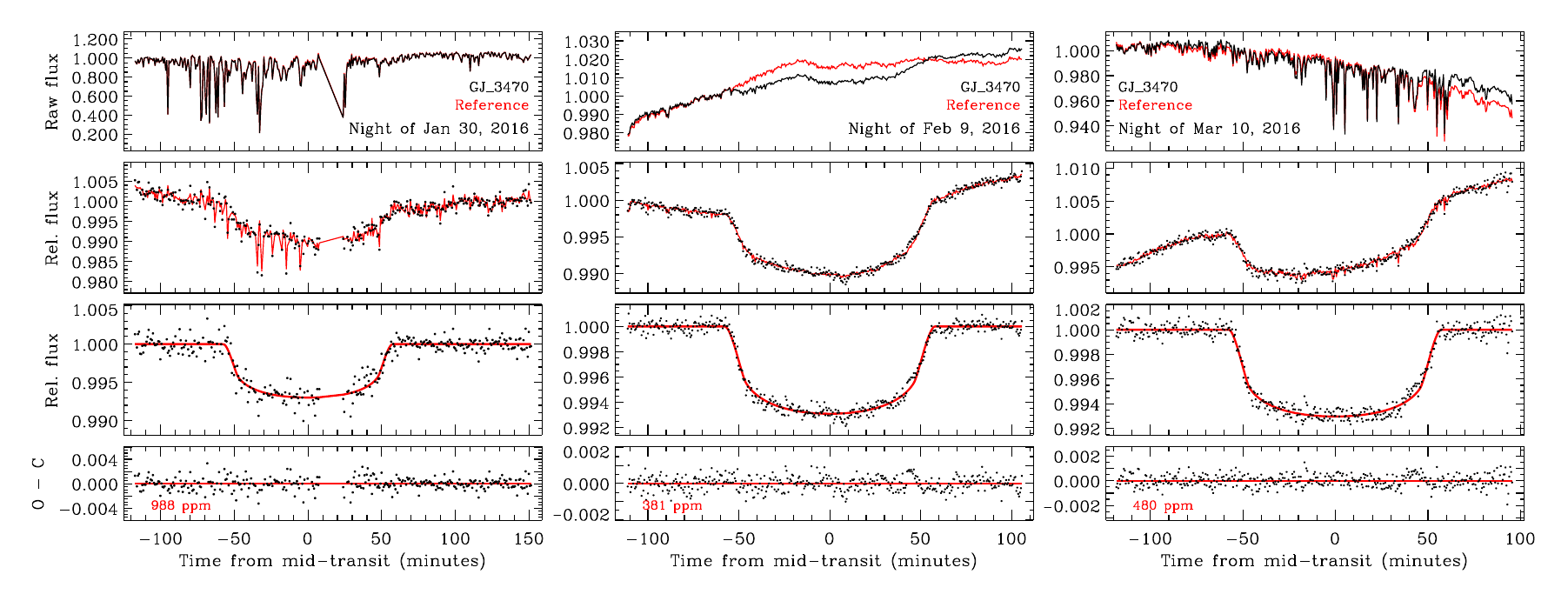}
 \caption{White-color light curves of \object{GJ 3470} obtained with GTC/OSIRIS on January 30 (Run 1; {\it left panel}), February 9 (Run 2; {\it middle panel}), and March 10 (Run 3; {\it right panel}) in 2016. In each panel, from top to bottom are: (1) raw flux time series of the target (black line) and reference (red line) stars; (2) raw light curve (target-to-reference flux ratio; black points) and the best-fitting combined model (red line); (3) corrected light curve after removing the best-fitting systematics model, overplotted with the best-fitting transit model; (4) best-fitting light-curve residuals. \label{fig:gtc_whiteLC}}
\end{figure*}

We first fit the white-color light curves to derive the overall transit parameters. For the GTC/OSIRIS observations, the rotator-angle dependent systematics have been a known issue in the previous studies when the rotator angle $\phi$ varies more than $\sim$60$^\circ$ and the stars are far away from the Nasmyth rotator center \citep[e.g.][]{2016A&A...594A..65N,chen2016a}. Such systematics will introduce sinusoid-like features that could distort the observed light curves. As summarized in Table~\ref{tab:osiris_obslog}, all three GTC/OSIRIS Runs spanned a rotator angle range larger than 90$^\circ$. The white-color light curves indeed exhibit sinusoid-like trends that cannot be solely described by the polynomial combinations of state vectors. Therefore, we modified the baseline model as: 
\begin{equation}
\mathcal{B}(c_j,c^\prime_k)=\mathcal{B}(c_j)\times \mathcal{B}^\prime(c^\prime_k),
\end{equation}
where $\mathcal{B}^\prime(c^\prime_k)=1+c^\prime_0\cos(c^\prime_1\phi+c^\prime_2)$ is the sinusoid function. 

To determine the optimal baseline model that can well decorrelate systematics, we tested all the polynomial combinations of different state vectors in linear to third-order forms. The state vectors included spectra's relative position shifts in the spatial and dispersing directions ($x$, $y$), spectra's full-width at the half maximum (FWHM) in the spatial and dispersing directions ($s_x$, $s_y$), airmass $z$, and time sequence $t$. We employed the Bayesian Information Criterion \citep[BIC;][]{Schwarz1978} to penalize the models that have too many degrees of freedom ($d.o.f.$). The final selected baseline models for the three Runs are listed below:
\begin{align}
\mathcal{B}_\mathrm{Run\_1} = & ~(c_0+c_1x+c_2y+c_3y^2+c_4s_y+c_5s_y^2+c_6z)\nonumber\\
& ~\times[1+c^\prime_0\cos(\phi+c^\prime_1)],\\
\mathcal{B}_\mathrm{Run\_2} = & ~(c_0+c_1y+c_2s_x+c_3s_y+c_4s_y^2+c_5z)\nonumber\\
& ~\times[1+c^\prime_0\cos(c^\prime_1\phi+c^\prime_2)],\\
\mathcal{B}_\mathrm{Run\_3} = & ~(c_0+c_1y+c_2s_x+c_3s_y+c_4s_x^2+c_5s_y^2+c_6t)\nonumber\\
& ~\times[1+c^\prime_0\cos(c^\prime_1\phi+c^\prime_2)].
\end{align}

We then performed three analyses for certain purposes:
\begin{itemize}
\item[--] Firstly, we jointly fit Runs 2 and 3 assuming that the free parameters ($i$, $a/R_\star$, $R_{\rm p}/R_\star$, $u_1$, $u_2$) were common. The mid-transit time $T_{\rm mid}$ and the coefficients ($c_j$, $c^\prime_k$) of the systematics baseline models were Run dependent. The derived transit parameters are given in Table~\ref{tab:gtc_param}, which are fully consistent with the ones obtained by the {\it Spitzer} observations in the 4.5~$\mu$m band and other literature studies \citep[see Table 2 of][]{2016MNRAS.463.2574A}. 
\item[--] Secondly, we fixed ($i$, $a/R_\star$, $T_{\rm mid}$) to the values listed in Table~\ref{tab:gtc_param} and fit $R_{\rm p}/R_\star$ individually. This results in $R_{\rm p}/R_\star$ values of 0.0770 $\pm$ 0.0024 and 0.0779 $\pm$ 0.0022 for Runs 2 and 3, respectively. The transit depth difference is well within the measured uncertainties. Although the impact of the associated instrumental systematics has been included in the large error bars, the best-fitting absolute transit depth value might be biased to some extent. Stellar activity could be another possibility (see the discussion in Sect.~\ref{sec:spotsonphot}).
\item[--] Thirdly, we derived the mid-transit time for Run 1. 
\end{itemize}

The white-color light curves of the three Runs are shown in Fig.~\ref{fig:gtc_whiteLC}. The standard deviation of the normalized residuals are 988 ppm (43~s cadence), 381 ppm (30~s cadence), and 480 ppm (30~s cadence) for Runs 1--3, respectively, which achieved 10.6, 2.4, and 3.1 times of the expected photon noise limits. Using the time-averaging $\beta$ approach \citep[e.g.][]{2006MNRAS.373..231P,2008ApJ...683.1076W,2010A&A...511A...3G}, the contribution of correlated noise was estimated as 255~ppm, 298~ppm, and 335~ppm for Runs 1--3, respectively.

\subsubsection{Revised transit ephemeris}

The 3 new mid-transit measurements from our GTC/OSIRIS observations are listed in Table~\ref{tab:gtc_param}. To refine the transit period, we also collected the mid-transit times of 13 transits reported in \citet{2016MNRAS.463.2574A} which included 3 re-analyzed transits of \citet{2012A&A...546A..27B}, 11 transits reported in \citet{2015ApJ...814..102D} which included 1 re-analyzed transit of \citet{2014MNRAS.443.1810B}, and 9 transits reported in \citet{2014MNRAS.443.1810B} which included 2 re-analyzed transits of \citet{2013A&A...559A..32N} and \citet{2013ApJ...770...95F}. The separately reported mid-transit times of \citet{2013A&A...551A..99C} and \citet{2013ApJ...768..154D} were also included. All the mid-transit times were fitted by the linear function: $T(E)=T_0+E\times P$ (see Table~\ref{tab:gtc_param}). The reduced chi-square for this fitting is $\chi^2_\mathrm{r}=1.23$, indicating that the linear ephemeris can well describe the current orbit. No significant transit time variation (TTV) is observed under current precision.

\begin{table}
     \small
     \centering
     \caption{Derived parameters for the GTC/OSIRIS observations}
     \label{tab:gtc_param}
     \begin{tabular}{cc}
     \hline\hline\noalign{\smallskip}
     Parameter & Value\\\noalign{\smallskip}
     \hline\noalign{\smallskip}
     \multicolumn{2}{c}{\it Transit Parameters from two-Run joint analysis\dotfill}\\\noalign{\smallskip}
     $R_{\rm p}/R_\star$ & 0.0777 $\pm$ 0.0026 \\\noalign{\smallskip}
     $i$ [degree] & 88.14 $^{+0.82}_{-0.64}$ \\\noalign{\smallskip}
     $a/R_\star$ & 13.20 $^{+0.86}_{-0.84}$ \\\noalign{\smallskip}
     $u_1$ & 0.430 $\pm$ 0.058 \\\noalign{\smallskip}
     $u_2$ & 0.387 $\pm$ 0.065 \\\noalign{\smallskip}
     \hline\noalign{\smallskip}
     \multicolumn{2}{c}{\it Mid-transit times\dotfill}\\\noalign{\smallskip}
     $T_{\rm mid}$ [$\mathrm{BJD}_\mathrm{TDB}$] & 2457418.46417 $\pm$ 0.00036 (Run 1) \\\noalign{\smallskip}
       & 2457428.47451 $\pm$ 0.00027 (Run 2) \\\noalign{\smallskip}
       & 2457458.50417 $\pm$ 0.00032 (Run 3) \\\noalign{\smallskip}
     \hline\noalign{\smallskip}
     \multicolumn{2}{c}{\it Revised ephemeris\dotfill}\\\noalign{\smallskip}
     $T_0$ [$\mathrm{BJD}_\mathrm{TDB}$] & 2455983.70417 $\pm$ 0.00011 \\\noalign{\smallskip}
     $P$ [days]   & 3.33665173 $\pm$ 0.00000059 \\\noalign{\smallskip}
    \hline\noalign{\smallskip}
    \end{tabular}
\end{table}

\subsubsection{Fitting of spectroscopic light curves}
\label{sec:fit_spect}

\begin{figure*}[h!]
 \centering
 \includegraphics[height=0.9\linewidth,angle=0.0]{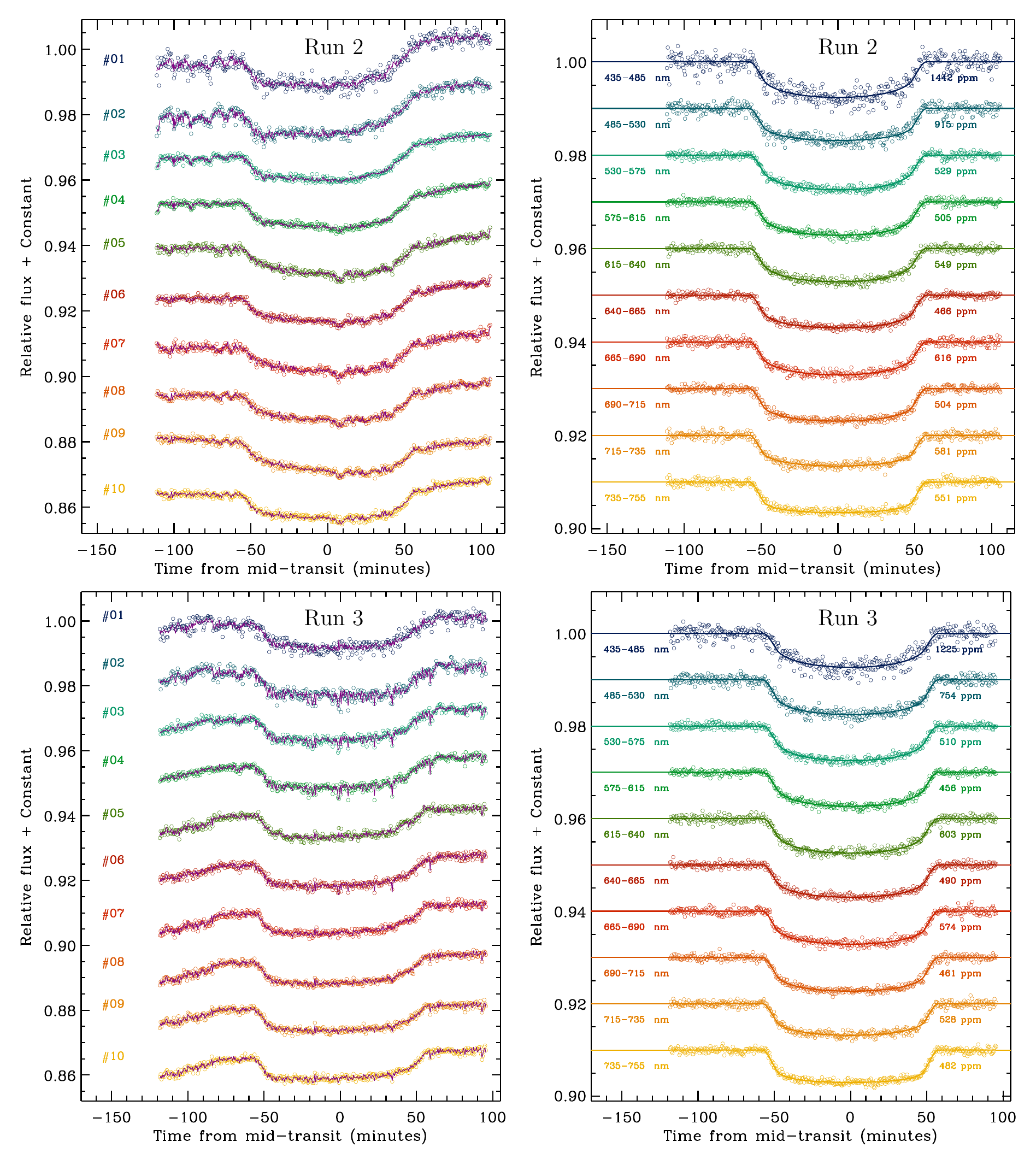}
 \caption{Raw ({\it left panels}) and detrended ({\it right panels}) spectroscopic light curves obtained with GTC/OSIRIS for Runs 2 and 3.\label{fig:gtc_specLC}}
\end{figure*}

To derive the transmission spectrum for \object{GJ 3470b}, we fit the spectroscopic light curves using the following systematics baseline model:
\begin{align}
\mathcal{B_\mathrm{spec}} = & ~\mathcal{S_\mathrm{w}}\times(c_0+c_1s_y+c_2s_y^2+c_3t+c_4t^2+c_5t^3),
\end{align}
where $\mathcal{S_\mathrm{w}}$ refers to the Run-dependent common-mode systematics derived from the white-color light curves. This model was also selected using the BIC values among a large set of polynomial combinations of state vectors which did not include any sinusoid trends. The common-mode systematics $\mathcal{S_\mathrm{w}}$ were derived after dividing the white-color light-curve data by the corresponding best-fitting transit model. According to the benchmark test on the transmission spectrum of a transiting white dwarf \citep{chen2016a}, the GTC/OSIRIS rotator-angle dependent systematics are achromatic, and the \texttt{divide-white} method that incorporates the common-mode trends in the baseline models could recover the buried transmission spectrum. 

\begin{figure}
 \centering
 \includegraphics[width=1.\linewidth,angle=0.0]{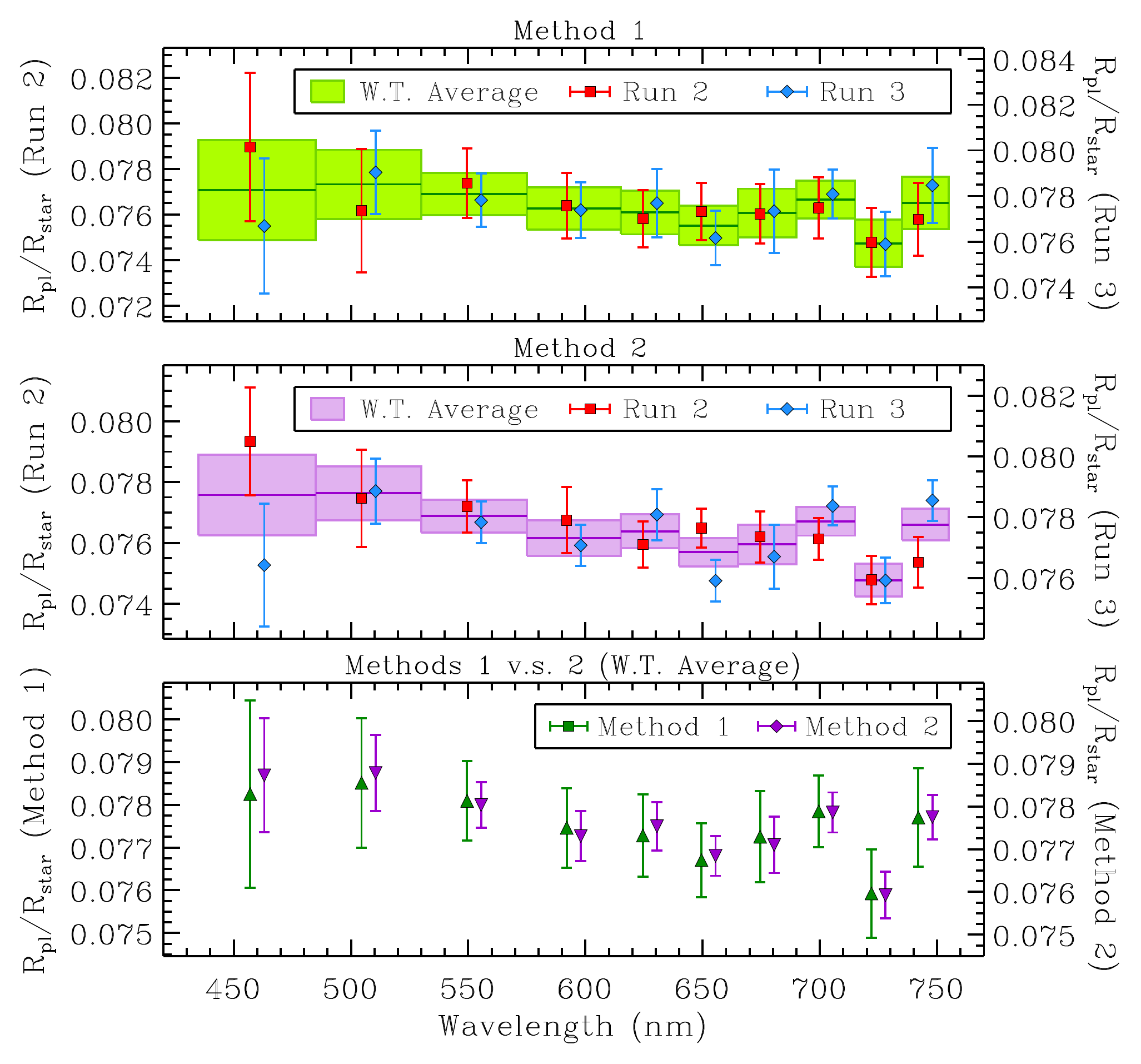}
 \caption{GTC/OSIRIS transmission spectra derived by different methods. Method 1 ({\it top panel}) refers to the Bayesian Information Criterion based model selection. Method 2 ({\it middle panel}) refers to the Akaike Information Criterion based systematics models marginalization. The single-Run transmission spectra are shown in red and blue for Runs 2 and 3, respectively. The shadowed boxes are the weighted average of the two nights, which are also separately shown in the {\it bottom panel}. The two data sets in each panel are shifted in wavelength for clarity. \label{fig:gtc_transpec_cmp}}
\end{figure}

We fixed the values of the common parameters ($i$, $a/R_\star$, $T_{\rm mid}$) to those obtained in the white-color light curves. The parameters ($R_{\rm p}/R_\star$, $u_1$, $u_2$) and baseline coefficients $c_j$ were fitted separately for different spectroscopic light curves. The resulting standard deviation of the normalized residuals for the spectroscopic light curves achieved a range of 1.0--1.9 and 1.0--1.7 times of the expected photon noise limits for Runs 2 and 3, respectively. The derived values of $R_{\rm p}/R_\star$ in different passbands are given in Table~\ref{tab:gtc_transpec}. The raw spectroscopic light curves and the detrended ones are shown in Fig.~\ref{fig:gtc_specLC}. The transmission spectra derived from Runs 2 and 3 are in full agreement with each other, with a chi-square value of $\chi^2=5.3$ ($d.o.f.=10$). If one is allowed to have an overall shift in transit depth to match another, the agreement becomes even better ($\chi^2=2.2$, $d.o.f.=9$). The resulting overall shift is $\Delta R_{\rm p}/R_\star=0.0012$, which is consistent with the difference in the white-color light-curve transit depths ($\Delta R_{\rm p}/R_\star=0.0009\pm0.0033$). We also note that the major deviation came from the bluest channel and the reddest channel while the in-between channels exhibited almost the same decreasing trend with increasing wavelengths (see the top panel of Fig.~\ref{fig:gtc_transpec_cmp}). 

\begin{table*}
     \footnotesize
     \centering
     \caption{Measured $R_{\rm p}/R_\star$ for the divided GTC/OSIRIS passbands}
     \label{tab:gtc_transpec}
     \begin{tabular}{cccccccc}
     \hline\hline\noalign{\smallskip}
     \# & $\lambda$ & \multicolumn{2}{c}{$R_{\rm p}/R_\star$ (Run \#2)} & \multicolumn{2}{c}{$R_{\rm p}/R_\star$ (Run \#3)} & \multicolumn{2}{c}{$R_{\rm p}/R_\star$ (Average)}\\\noalign{\smallskip}
        & (nm)  & Method 1 & Method 2 & Method 1 & Method 2 & Method 1 & Method 2\\\noalign{\smallskip}
     \hline\noalign{\smallskip}
1  & 435--485 & 0.0790 $\pm$ 0.0033 & 0.0793 $\pm$ 0.0018 & 0.0767 $\pm$ 0.0030 & 0.0764 $\pm$ 0.0020 & 0.0783 $\pm$ 0.0022 & 0.0787 $\pm$ 0.0013\\\noalign{\smallskip} 
2  & 485--530 & 0.0762 $\pm$ 0.0027 & 0.0775 $\pm$ 0.0016 & 0.0790 $\pm$ 0.0018 & 0.0789 $\pm$ 0.0011 & 0.0785 $\pm$ 0.0015 & 0.0788 $\pm$ 0.0009\\\noalign{\smallskip} 
3  & 530--575 & 0.0774 $\pm$ 0.0015 & 0.0772 $\pm$ 0.0009 & 0.0778 $\pm$ 0.0012 & 0.0778 $\pm$ 0.0007 & 0.0781 $\pm$ 0.0009 & 0.0780 $\pm$ 0.0005\\\noalign{\smallskip} 
4  & 575--615 & 0.0764 $\pm$ 0.0014 & 0.0767 $\pm$ 0.0011 & 0.0774 $\pm$ 0.0012 & 0.0771 $\pm$ 0.0007 & 0.0775 $\pm$ 0.0009 & 0.0773 $\pm$ 0.0006\\\noalign{\smallskip} 
5  & 615--640 & 0.0758 $\pm$ 0.0012 & 0.0759 $\pm$ 0.0008 & 0.0777 $\pm$ 0.0015 & 0.0781 $\pm$ 0.0008 & 0.0773 $\pm$ 0.0010 & 0.0775 $\pm$ 0.0006\\\noalign{\smallskip} 
6  & 640--665 & 0.0761 $\pm$ 0.0013 & 0.0765 $\pm$ 0.0006 & 0.0762 $\pm$ 0.0012 & 0.0759 $\pm$ 0.0007 & 0.0767 $\pm$ 0.0009 & 0.0768 $\pm$ 0.0005\\\noalign{\smallskip} 
7  & 665--690 & 0.0760 $\pm$ 0.0013 & 0.0762 $\pm$ 0.0008 & 0.0773 $\pm$ 0.0018 & 0.0767 $\pm$ 0.0011 & 0.0773 $\pm$ 0.0011 & 0.0771 $\pm$ 0.0007\\\noalign{\smallskip} 
8  & 690--715 & 0.0763 $\pm$ 0.0013 & 0.0761 $\pm$ 0.0007 & 0.0781 $\pm$ 0.0011 & 0.0784 $\pm$ 0.0006 & 0.0778 $\pm$ 0.0008 & 0.0779 $\pm$ 0.0005\\\noalign{\smallskip} 
9  & 715--735 & 0.0748 $\pm$ 0.0015 & 0.0748 $\pm$ 0.0008 & 0.0759 $\pm$ 0.0014 & 0.0759 $\pm$ 0.0007 & 0.0759 $\pm$ 0.0010 & 0.0759 $\pm$ 0.0005\\\noalign{\smallskip} 
10 & 735--755 & 0.0758 $\pm$ 0.0016 & 0.0754 $\pm$ 0.0008 & 0.0785 $\pm$ 0.0016 & 0.0785 $\pm$ 0.0007 & 0.0777 $\pm$ 0.0011 & 0.0778 $\pm$ 0.0005\\\noalign{\smallskip} 
    \hline\noalign{\smallskip}
    \end{tabular}
    \tablefoot{Methods 1 and 2 refer to the BIC-based model selection and the AIC-based marginalization, respectively. For Method 1, the weighted average is calculated after shifting Run \#2 upwards $\Delta R_{\rm p}/R_\star=0.00119$. For Method 2, the weighted average is calculated after shifting Run \#2 upwards $\Delta R_{\rm p}/R_\star=0.00115$.}
\end{table*}

\citet{2014MNRAS.445.3401G} proposed that marginalization over many systematics is more robust than simple model selection, and that the BIC-based model selection could be the worst criterion in their experiments. To assess the impact of BIC-based model selection choices (hereafter method 1) on our derived transmission spectrum, we also performed a separate analysis on the spectroscopic light curves employing the systematics marginalization approach (hereafter method 2). We followed the implication of this approach described in \citet{2016ApJ...819...10W}, and refer the reader to that work for more details. Instead of using the wavelet-based MCMC to account for the correlated noise (see Method 1), for simplicity, Method 2 employed the \texttt{MPFIT} package to fit the data and accounted for the correlated noise using the time-averaging $\beta$ approach. We calculated the marginal likelihood for all the systematics models using the Akaike Information Criterion \citep[AIC;][]{akaike1973} as the approximation, i.e. $\ln\mathcal{P}(D|S_q)\approx-\mathrm{AIC}/2$, which provides more adequate fits and performs better than BIC as suggested by \citet{2014MNRAS.445.3401G}. The resulting $R_{\rm p}/R_\star$ in each spectroscopic channel was then calculated as the marginal-likelihood-weighted average of the best-fitting values from all the systematics models, whose uncertainty was propagated from both the deviation from the weighted average and the best-fitting error bar for each systematics model. The derived $R_{\rm p}/R_\star$ values are also listed in Table~\ref{tab:gtc_transpec}. The middle and bottom panels of Fig.~\ref{fig:gtc_transpec_cmp} show the transmission spectra derived by Method 2 and the comparison between these two methods, respectively. The great consistency confirms that the BIC-based model selection in this work does not bias the derived transmission spectrum. Since the two methods have almost the same transit-depth values in any given spectral channel, we decide to present the results from Method 2 in the following discussion, as they have smaller error bars.

\subsubsection{Re-analysis of Nascimbeni et al.'s LBT light curves}

For the consistency of our work, we re-analyzed the two LBT transit light curves from \citet{2013A&A...559A..32N} following the Method 1, with a simple baseline model $\mathcal{B} = c_0+c_1t$. We fixed the values of $P$, $i$, $a/R_\star$ to those listed in Table~\ref{tab:gtc_param}, and treated the limb-darkening coefficients in the same way as described in Sect.~\ref{sec:general_analysis}. The derived planet-to-star radius ratio is $0.0805\pm 0.0021$ in the $U_{\rm spec}$ filter and $0.07488\pm 0.00057$ in the $F972N20$ filter. We will interpret this re-analyzed data together with our GTC/OSIRIS data in Sect.~\ref{sectV}.

\section{High resolution spectra}
\label{sectIII}

\begin{figure*}[!ht]
 \centering
 \includegraphics[width=1\linewidth,angle=0.0]{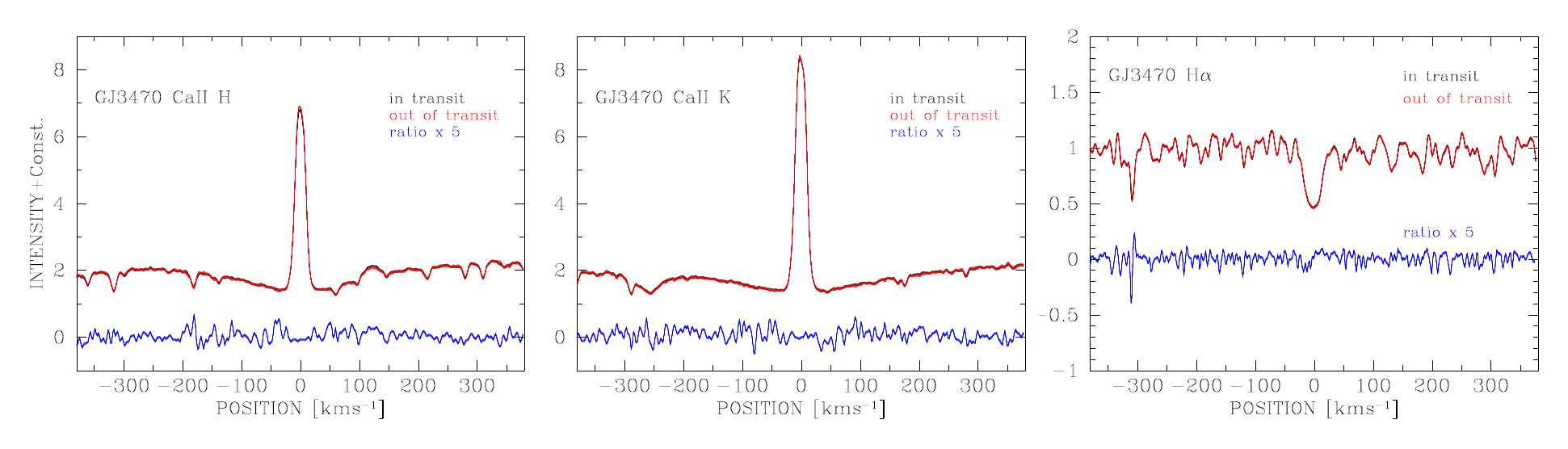}
 \caption{\ion{Ca}{II}\,H ({\it left}), \ion{Ca}{II}\,K ({\it middle}) and H$\alpha$ ({\it right}) of \object{GJ 3470} in- (black) and out-of-transit (red). The blue line is the ratio between the two. \label{fig:vlt_ratio_spec}}
\end{figure*}

\subsection{Observations and data reduction}

In order to assess the impact of stellar activity on transit observations and to search for narrow spectral signatures in the transmission spectrum, a transit of GJ 3470b was observed with the high resolution Ultraviolet and Visual Echelle Spectrograph \citep[UVES;][]{2000SPIE.4008..534D} at the ESO's Very Large Telescope (VLT) in program 096.C-0258(A). The observation was conducted continuously on January 14, 2016 from 04:50 UT to 08:40 UT (JD 2457401.701 to JD 2457401.861). The flat part of the transit lasted from 05:57 UT to 07:30 UT and the whole transit from 05:47 UT to 07:41 UT. During this time 22 spectra with exposure times of 568s were obtained. The spectra cover the wavelength range from 325.9 to 449.3 nm in the blue and 472.6 nm to 683.5 nm in the red part. Due to an earth-quake on November 27, 2015 of magnitude 6.2, at a distance of $\sim$30\,km from Paranal, the resolution of UVES in the blue channel was slightly reduced to $\lambda/\Delta \lambda = 45000$ instead of $\lambda/\Delta \lambda = 55000$ with the 0.8 arcsec slit used. The resolution in the red channel is $\lambda/\Delta \lambda = 52000$. The standard reduction pipeline was used for the data reduction, but the raw spectra were also manually reduced using standard IRAF tools with no effect on the scientific results.

\subsection{Results}
\label{sec:hiresresults}

\subsubsection{Change of line flux between in- and out-of-transit}
\label{sec:lineflux}

We derived a flux of $(0.95\pm0.002)\times10^{-14}$ $\rm ergs\,cm^{-2}\,sec^{-1}$, and $(1.22\pm0.008)\times10^{-14}$ $\rm ergs\,cm^{-2}\,sec^{-1}$ for the \ion{Ca}{II}\,H and K lines, respectively. With a distance of $28.8\pm2.5$~pc \citep{2014MNRAS.443.1810B}, this gives $\log (\ion{Ca}{II}$\,H+K) = $5.15\pm0.21$ (cgs). 

Fig.\,\ref{fig:vlt_ratio_spec} shows the lines of \ion{Ca}{II}\,H+K and H$\alpha$ taken in- and out-of-transit. The difference is so small that it can hardly be seen. The blue line below shows the ratio in- to the out-of-transit spectrum. The relative flux of the \ion{Ca}{II}\,H+K emission cores changes by only $0.67\pm0.22$\%. Because M-stars are very faint in the UV, the H$\alpha$ line is often used as activity indicator. Again, there is no significant difference seen in the in- and out-of-transit spectrum. The relative flux change is even smaller for H$\alpha$, only $0.21\pm0.19\%$, and thus no significant change of the line flux, or equivalent width, are found.

\subsubsection{High-resolution transmission spectra}

\begin{figure}
 \centering
 \includegraphics[height=0.85\linewidth,angle=-90.0]{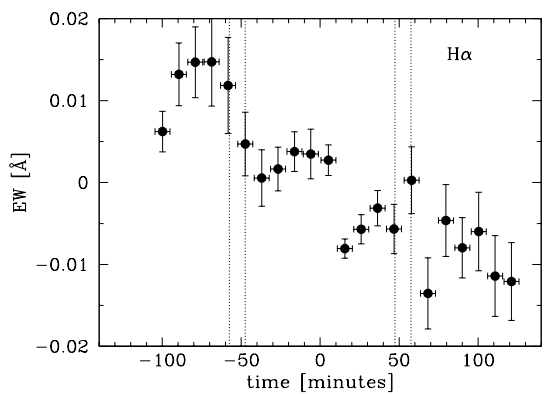}
 \caption{Equivalent width of the H$\alpha$ transmission spectrum. The dashed lines indicate the beginning and the end of the transit,\label{fig:vlt_ha_transpec}}
\end{figure} 

As pointed out by \citet{2016AJ....152...20C} that \ion{Ca}{II}\,H+K lines can be used to find out if any pre-transit Balmer line signal is caused by stellar activity or by material escaping from the planet. The transmission spectrum of planetary atmosphere is defined as: 
\begin{align}
S_T &= {{F_\mathrm{in}}\over{F_\mathrm{out}}}-1,
\end{align}
and the equivalent width of the transmission spectrum is defined as the integral from $-200$~km~s$^{-1}$ to $+200$~km~s$^{-1}$:
\begin{align}
W_\lambda &= \sum^{+200}_{\nu=-200}\Bigg(1-\frac{F_\nu}{F_\nu^\mathrm{out}}\Bigg)\,\Delta\lambda_\nu.
\end{align}

In order to study the planetary atmosphere or the material that might be escaping from the planet, we derived the transmission spectrum in the \ion{Na}{I}\,$\rm D_1$, \ion{Na}{I}\,$\rm D_2$, H$\alpha$, and tried to find out if there are effects from the stellar activity in \ion{Ca}{II}\,H+K. The equivalent width variations of H$\alpha$ are shown in Fig.\,\ref{fig:vlt_ha_transpec}. The planetary atmosphere would cause an absorption feature during the transit that is not seen before or after the transit. However, we observed a constant decrease in H$\alpha$ and an increase in the \ion{Ca}{II}\,H+K lines. The equivalent widths of the \ion{Ca}{II}\,H+K and H$\alpha$ transmission spectra are well correlated, with a correlation factor of $-0.85$. Given that H$\alpha$ is an absorption line and \ion{Ca}{II}\,H+K are emission lines, the best explanation is that the change of the equivalent width is caused by a small decrease of the active regions that are visible on the stellar surface. However, this effect is tinny (see Table~\ref{tab:ew}) compared to those of the stellar atmosphere, as stellar H$\alpha$ has an equivalent width of $0.411\pm 0.017$~\AA, stellar \ion{Ca}{II}\,H has $-3.558\pm 0.048$~\AA, and stellar \ion{Ca}{II}\,K has $-5.432\pm 0.063$~\AA. Thus the change of the active regions' size is quite small and can be explained simply by active regions that rotate out of view. There is no significant change of the Na lines during the transit. We thus conclude that we did not detect the planetary atmosphere in the Balmer, \ion{Ca}{II}, and Na lines. 

\begin{table}
     \small
     \centering
     \caption{Equivalent widths of the H$\alpha$, \ion{Na}{I} and \ion{Ca}{II} transmission spectra}
     \label{tab:ew}
     \begin{tabular}{cccc}
     \hline\hline\noalign{\smallskip}
     Spectral line & \multicolumn{3}{c}{Equivalent width $W_\lambda$ ($10^{-3}$~\AA)}\\\noalign{\smallskip}
     & Pre-transit & In-transit & Post-transit\\\noalign{\smallskip}
     \hline\noalign{\smallskip}
     H$\alpha$ & $8.50\pm 0.98$ & $-0.40\pm 0.99$ & $-5.53\pm 1.20$ \\\noalign{\smallskip}
     \ion{Na}{I}\,D$_1$+D$_2$ & $-0.37\pm 0.53$ & $0.41\pm 0.43$ & $-0.33\pm 0.76$ \\\noalign{\smallskip}
     \ion{Ca}{II}\,H+K & $-77.13\pm 6.42$ & $-8.14\pm 8.48$ & $56.84\pm 16.20$ \\\noalign{\smallskip}
    \hline\noalign{\smallskip}
    \end{tabular}
\end{table}

\section{Stellar activity}
\label{sectIV}

\subsection{Chromospheric activity}
\label{sec:plage}

The plage regions are characterized by the chromospheric emission cores in the resonance lines like \ion{Ca}{II}\,H+K and \ion{Mg}{II}\,h+k. With the emission flux of \ion{Ca}{II}\,H+K lines obtained in Sect.~\ref{sec:lineflux}, we can derive the magnetic filling factor, which is defined as the fraction of the stellar surface covered by magnetic fields. 

\citet{2011MNRAS.414.2629M} determined the fluxes in the \ion{Ca}{II}\,H+K lines for magnetically saturated stars using high-resolution spectra. They derived $\log(\ion{Ca}{II}$\,H+K) values of $6.01\pm 0.07$~(cgs) for V105 Or (M1.5V) and $6.00\pm 0.03$~(cgs) for V371 Ori (M2V). Since the magnetic filling factor of these stars has to be no more than 100\%, this gives us an upper limit of the magnetic filling factor of $\sim$15\% for \object{GJ 3470}. \citet{2002A&A...386..983F} calculated the heating of the chromosphere and the fluxes in the \ion{Ca}{II}\,H+K lines for stars of different spectral type theoretically. By rescaling these values to the observed fluxes in the \ion{Ca}{II}\,H+K lines, they derived a relation between the fluxes of these lines and the magnetic filling factor. If we use the fluxes obtained with high-resolution spectra as published by \citet{2011MNRAS.414.2629M} and this model, we can roughly estimate that the magnetic filling factor of \object{GJ 3470} to be larger than $\sim$10\%. The \ion{Ca}{II}\,H+K filling factor on the Sun varies typically within 1--9\% from the solar minimum to the maximum \citep{2009A&A...501.1103M}. In this sense, with a filling factor of 10--15\%, \object{GJ 3470} is more active than the Sun. 

Given that the planet occults only 0.6\% of the stellar surface, this large filling factor means that it is impossible for the planet to occult all the plage regions at any one moment. Therefore, only when the plage regions are disproportionally occulted, i.e. if they are preferentially located along the path of the planet, we would see a significant decrease of the emission line flux of \ion{Ca}{II}\,H+K. Since it only changes by $0.67\pm0.22$\%, the plage regions are likely very homogeneously distributed over the stellar surface.

\subsection{Photospheric activity}
\label{sec:spotsonphot}

\begin{figure}
 \centering
 \includegraphics[width=1\linewidth,angle=0.0]{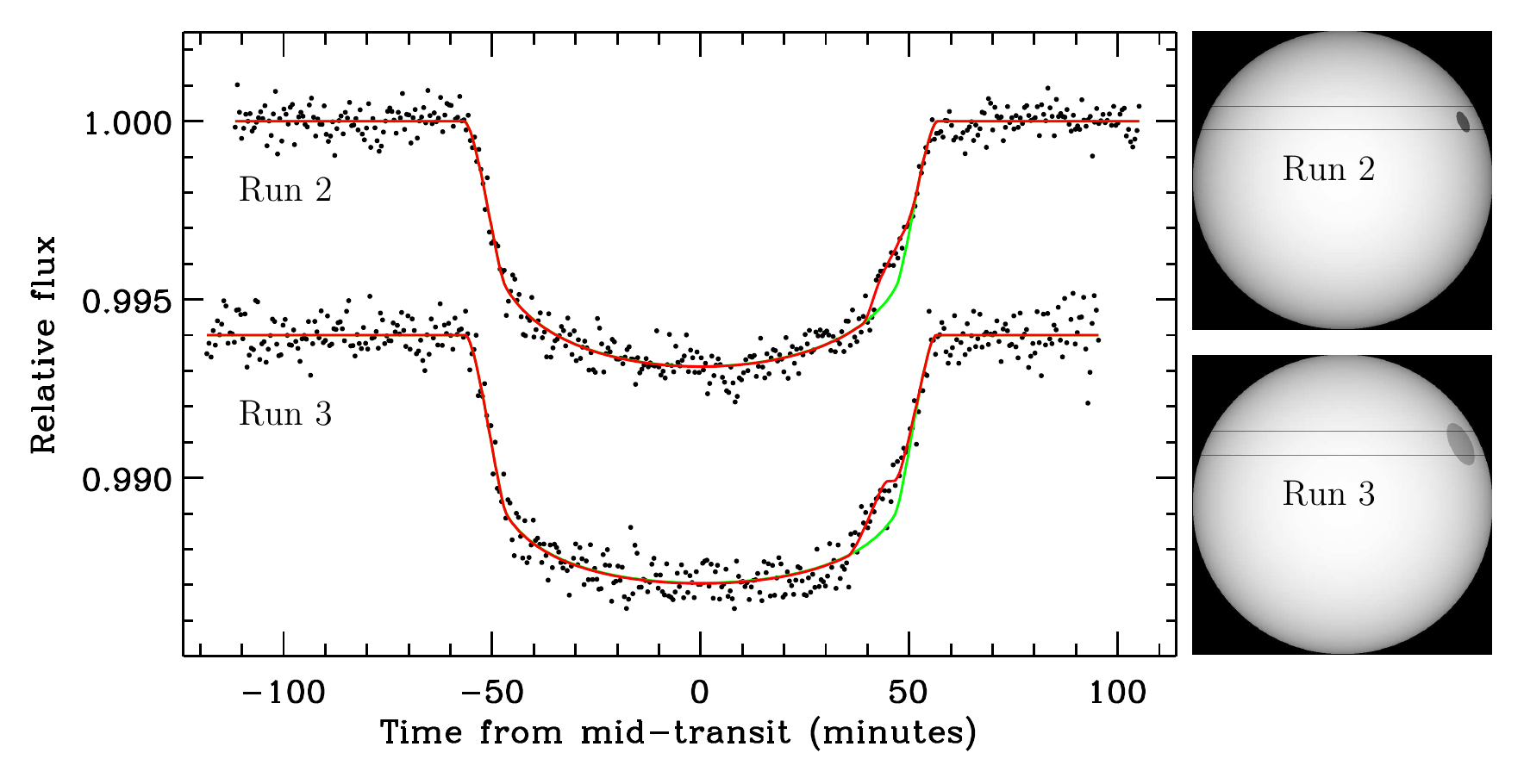}
 \caption{Modeling the light curves with the assumption of occulted spots. The red and green lines show the spotted transit model and non-spotted transit model, respectively.\label{fig:spottedlc}}
\end{figure}

While high-resolution spectroscopy can probe active regions in the chromosphere (e.g. plages), white-color light curves can provide information of star spots that are originating from the photosphere. Our GTC/OSIRIS white-color light curves showed a small offset in planet-to-star ratios between Runs 2 (0.0770 $\pm$ 0.0024) and 3 (0.0779 $\pm$ 0.0022), which could be caused by either observational systematics or the impact of stellar activity. Due to the normalization effect in the light-curve modeling process, the apparent transit depth would change from epoch to epoch due to the modulation of star spots. According to the photometric monitoring by \citet{2013ApJ...770...95F} and \citet{2014MNRAS.443.1810B}, \object{GJ 3470} did show a peak-to-valley amplitude of $\sim$1\% with a rotational period of $20.7\pm 0.15$~days.

Given that the bump-like time-correlated noise could come from occulted dark star spots, we re-modeled the GTC/OSIRIS light curves with the \texttt{PRISM+GEMC} code \citep{2013MNRAS.428.3671T,2015MNRAS.450.1760T}, which uses a pixellation approach in Cartesian coordinates to generate the spotted transit light curve (see Fig.~\ref{fig:spottedlc}). The fitted spot angular radius and contrast were $4.3^\circ\pm 11.8^\circ$ and $0.45\pm 0.23$ for Run 2, and $8.9^\circ\pm 10.1^\circ$ and $0.75\pm 0.44$ for Run 3. Using the PHOENIX stellar atmosphere models and GTC/OSIRIS response, we found that the spots were $366^{+175}_{-267}$~K and $139^{+233}_{-370}$~K cooler than the photosphere ($T_{\rm spot}=3652$~K), respectively. The spot-photosphere temperature differences are in line with the decreasing trend as a function of decreasing photosphere temperature \citep[e.g.][]{2005LRSP....2....8B,2015MNRAS.448.3053A}. Assuming that the stellar variability is also 1\% in our GTC/OSIRIS passband and that any unocculted star spots have similar contrast, the spot filling factors would be $1.8\pm 0.8$\% and $4\pm 7$\% for Runs 2 and 3, which are clearly smaller than the magnetic filling factor (10--15\%). The spot filling factor has been found to be strongly deviating from and smaller than the magnetic filling factor for cool late-type stars, suggesting that they refer to different activity signatures \citep{2005LRSP....2....8B}.

These time-correlated noise could in principle also be modeled by occulted bright spots. In that case, a minimum spot filling factor of $36\pm 14$\% and $49\pm 10$\% is derived for Runs 2 and 3, respectively. The fitted contrasts of $1.33\pm 0.11$ and $1.23\pm 0.07$ indicate that these spots are $154\pm 46$~K and $109\pm 32$~K hotter than the photosphere, respectively. This would translate into a flux variability larger than $\sim$12\%, which strongly disagrees with the long-term monitoring photometry. Furthermore, introducing occulted bright spots would make the real transit depth much shallower and more discrepant from the infrared transit observations. Therefore, the scenario of occulted bright spots is unlikely the cause of the correlated noise.

\subsection{Impact of stellar activity on transmission spectrum}
\label{sec:activityimpact}

\begin{figure}
 \centering
 \includegraphics[width=0.9\linewidth,angle=0.0]{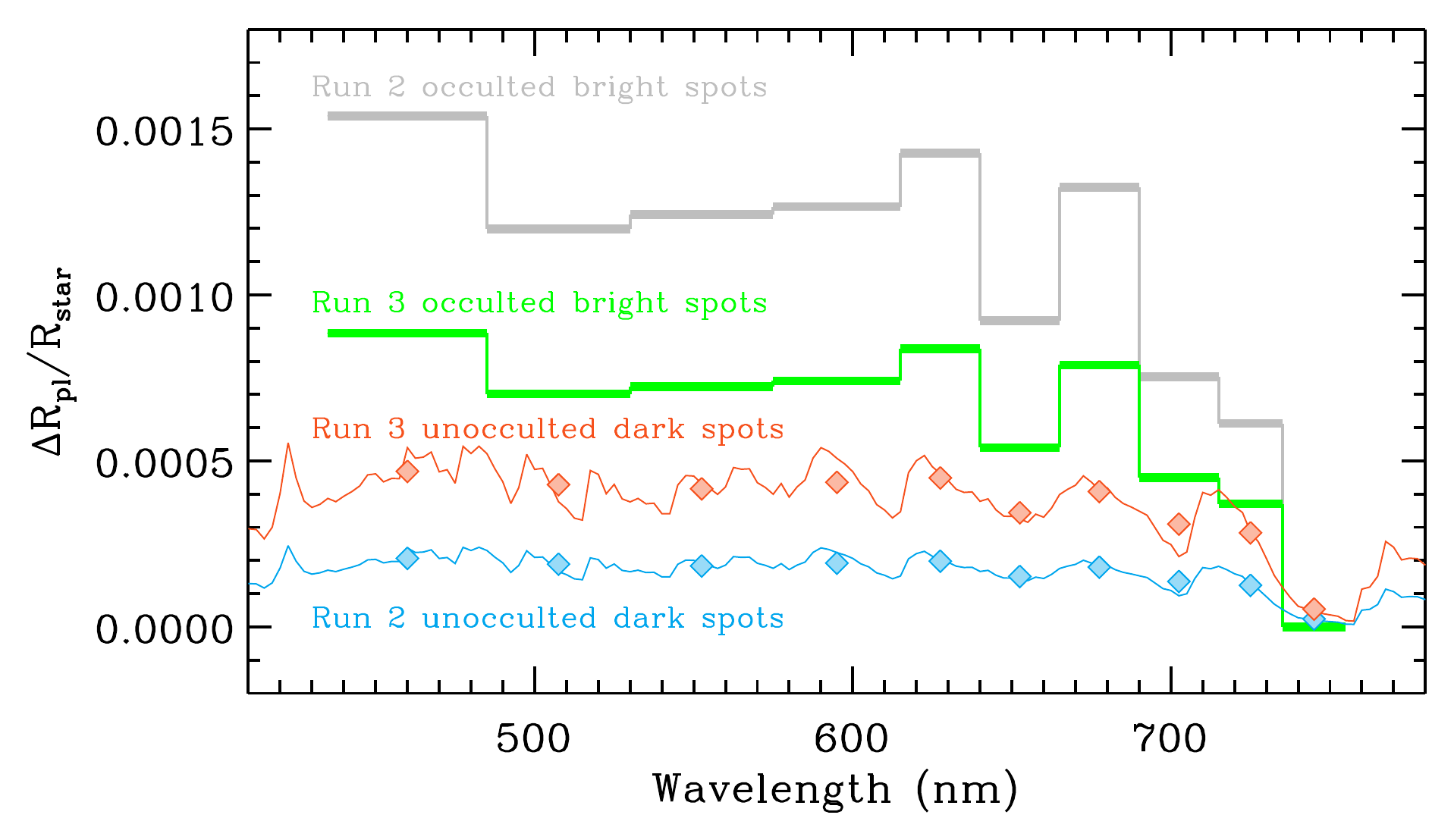}
 \caption{The impact of stellar activity on a flat constant line. These apparent relative transmission spectra were calculated using the fitting results obtained in Sect.~\ref{sec:spotsonphot} for the cases of unocculted dark spots (lines with diamonds) and occulted bright spots (histogram lines), respectively. \label{fig:spotimpact}}
\end{figure}

We now attempt to assess the impact of stellar activity on the derived transmission spectrum in a single epoch. The occulted dark star spots discussed in Sect.~\ref{sec:spotsonphot}, if they were real, exhibited non-detectable wavelength dependent variation, thus having negligible impact on the transmission spectrum. 

For the impact of unocculted dark star spots, we calculated the apparent transit depths $\tilde{R}^2_\mathrm{p}/\tilde{R}^2_\star$ using the Equation (1) presented in \citet{2014ApJ...791...55M}:
\begin{align}
\frac{\tilde{R}^2_\mathrm{p}}{\tilde{R}^2_\star} &= \frac{R^2_\mathrm{p}}{R^2_\star}\frac{1}{1-\delta[1-F_\nu({\mathrm{spot}})/F_\nu({\mathrm{phot}})]},
\end{align}
where $F_\nu({\mathrm{spot}})$ and $F_\nu({\mathrm{phot}})$ are the spectra of the spots and the photosphere, respectively, and $\delta$ is the spot filling factor. We adopted the PHOENIX stellar atmosphere models, and used the spot filling factors and spot temperatures determined in Sect.~\ref{sec:spotsonphot}. As shown in Fig.~\ref{fig:spotimpact}, if there were unocculted dark star spots, a flat spectrum at the constant value of $R_{\rm p}/R_\star=0.0777$ would appear to be wavelength dependent in the measurements. Due to strong absorption bands of TiO/VO in the cool stellar atmosphere, the apparent transmission spectrum binned in our passbands is relatively flat in the wavelengths 435--735~nm rather than mimicking a Rayleigh scattering slope, where the maximum differences are only $\Delta R_{\rm p}/R_\star=0.00008$ and 0.00018 for Runs 2 and 3, respectively. The apparent planetary radius is much smaller in the 735--755~nm band, making the relative spectral shape more deviating from a potential Rayleigh scattering slope and our measurements. 

Since the maximum difference of the apparent planetary radius in the GTC/OSIRIS passband is much smaller than the average radius ratio uncertainty, this impact plays a negligible role in changing the shape of the observed transmission spectrum. Therefore, we conclude that (i) the variation in the transmission spectrum of our single Run, and (ii) the difference between the transmission spectra of our two Runs, are not likely caused by the unocculted dark star spots. However, it should be noted that the spot spectrum does not necessarily follow the same as the photosphere. \citet{2011MNRAS.416.1443S} analyzed the wavelength dependent effect of the occulted star spots in HD 189733b, and found that the expected MgH feature at the wavelength 500~nm was weaker than expected. In our case, it is still possible that an ad hoc spot spectrum happens to cancel out all the TiO/VO bands to show a Rayleigh scattering like slope. But this is beyond the scope of this discussion.

\citet{2014A&A...568A..99O} has attempted to use occulted bright spots to explain the transmission spectrum of \object{GJ 3470b}. This worked well if blackbody emission is assumed for both bright spots and photosphere. Here we again adopted the PHOENIX stellar atmosphere models, and simply assumed that the bright spots were 100~K hotter than the photosphere. The location and size of the occulted bright spots were inherited from the fitting results determined in Sect.~\ref{sec:spotsonphot}. Synthetic spotted light curves were created using the spot-to-photosphere flux ratios in the corresponding passbands, and then were fitted by the non-spotted transit model to derive the final apparent transmission spectrum. As shown in Fig.~\ref{fig:spotimpact}, the apparent transmission spectrum resulting from occulted bright spots would introduce similar relative variation to that from the unocculted dark spots, but in a larger scale. The same conclusion as the unocculted dark spots also applies to the occulted bright spots. On the other hand, according to the discussion in Sect.~\ref{sec:spotsonphot}, we also emphasize that the occulted bright spots are not likely present in our observed light curves.

\section{Interpreting the transmission spectrum as planetary atmosphere}
\label{sectV}

For the discussion in this section, we calculated a weighted average GTC/OSIRIS transmission spectrum after shifting upwards $\Delta R_{\rm p}/R_\star=0.00115$ to let Run 2 match Run 3. The change of the relative shape of transmission spectrum caused by stellar activity is assumed to be negligible. The average transmission spectrum is also presented in Table~\ref{tab:gtc_transpec}.

\subsection{Spectrally-resolved scattering slope}

\begin{figure}
 \centering
 \includegraphics[width=\linewidth,angle=0.0]{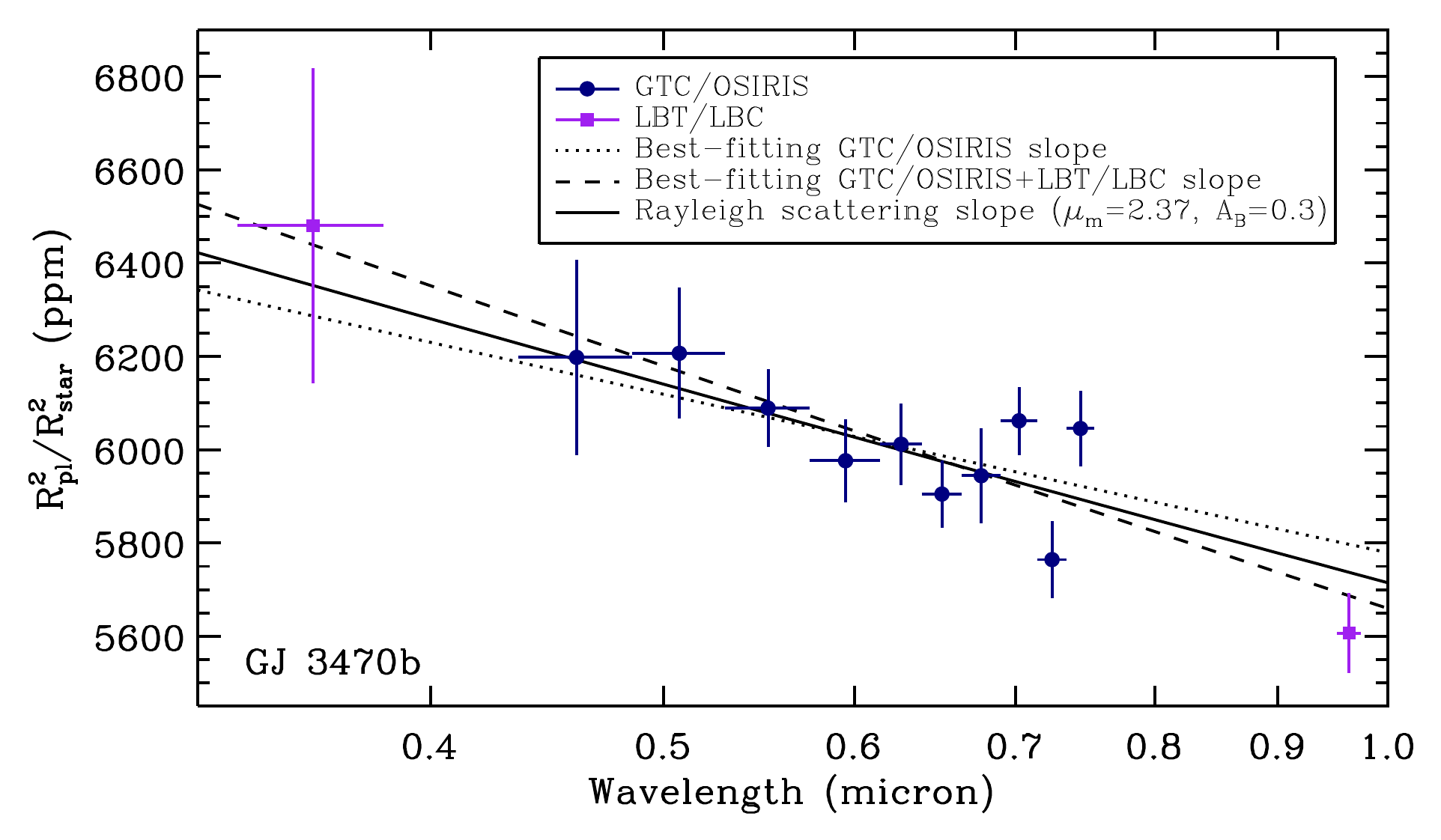}
 \caption{Optical transmission spectrum of \object{GJ 3470b}. The GTC/OSIRIS measurements and re-analyzed LBT/LBC measurements are shown in navy-blue circles and purple squares with error bars, respectively. The dotted line shows the best-fitting slope of the GTC/OSIRIS measurements. The dashed line shows the best-fitting slope of the combined GTC/OSIRIS and LBT/LBC measurements. The solid line shows a Rayleigh scattering slope based on the physical parameters of \object{GJ 3470b} with a mean molecular weight of $\mu_{\rm m}=2.37$ and a Bond albedo of $A_{\rm B}=0.3$. \label{fig:transpec_opt}}
\end{figure}

As shown in Fig.~\ref{fig:transpec_opt}, our derived GTC/OSIRIS transmission spectrum shows a clear trend of decreasing planetary radii with increasing wavelengths. Therefore, we performed a linear fit to this trend in the ($\ln\lambda$, $R_{\rm p}/R_\star$) space, and derived a slope of $-$0.0032 $\pm$ 0.0016 ($\chi^2=11.2$; $d.o.f.$=8). According to \citet{2008A&A...481L..83L}, if the opacity sources in the planetary atmosphere have the cross section in the form of $\sigma=\sigma_0(\lambda/\lambda_0)^\alpha$, a slope would be introduced into the transmission spectrum, which is connected to the atmospheric scale height $H$:
\begin{align}
\frac{\mathrm{d}\,R_\mathrm{p}}{\mathrm{d}\ln\lambda} &= \alpha H = \alpha\frac{k_\mathrm{B}T_\mathrm{p}}{\mu_\mathrm{m} g_\mathrm{p}},
\end{align}
where $k_{\rm B}$ is the Boltzmann constant, $T_{\rm p}$ and $\mu_\mathrm{m}$ are the temperature and mean molecular weight of the planetary atmosphere, respectively, and $g_{\rm p}$ is the planetary surface gravity. For \object{GJ 3470b}, we obtained $g_{\rm p}=6.8\pm 1.7$~m\,s$^{-2}$, $R_\star=0.48\pm 0.04\,R_\sun$ and $T_{\rm eff}=3652\pm 50$~K from \citet{2014MNRAS.443.1810B}, and calculated a new equilibrium temperature as $T_{\rm p}=(1-A_{\rm B})^{1/4}\times(711\pm 25)$~K. The Bond albedo $A_{\rm B}$ could be zero for a dark surface in the extreme case. The typical value for Uranus and Neptune\footnote{See \url{http://nssdc.gsfc.nasa.gov/planetary/factsheet/}.} in our Solar system is 0.3. 

According to our fitting result, if \object{GJ 3470b}'s atmosphere is composed of solar-composition H$_2$/He dominated atmosphere (i.e. $\mu_\mathrm{m}=2.37$), the slope in the cross section would be $\alpha=-2.9\pm1.6$ ($A_{\rm B}$=0.0) or $\alpha=-3.2\pm1.8$ ($A_{\rm B}$=0.3), which covers the Rayleigh scattering slope $\alpha=-4$ within the error bar. On the other hand, if Rayleigh scattering is assumed, a mean molecular weight of $\mu_\mathrm{m}=3.3\pm 1.8$ ($A_{\rm B}$=0.0) or $\mu_\mathrm{m}=2.9\pm 1.7$ ($A_{\rm B}$=0.3) would be required for the atmosphere, which is slightly heavier than the solar-composition H$_2$/He atmosphere. 

\citet{2013A&A...559A..32N}'s LBT observations were conducted in two filters either bluer or redder than our GTC/OSIRIS wavelength range. Neglecting the impact of stellar activity, if the two re-analyzed LBT measurements are included in our aforementioned fitting, we would derive a slope of $-0.0049\pm 0.0010$. This corresponds to a cross-section slope of $\alpha$ = $-4.5\pm 1.5$ ($A_{\rm B}$=0.0, $\mu_\mathrm{m}$=2.37) or $\alpha$ = $-4.9\pm 1.7$ ($A_{\rm B}$=0.3, $\mu_\mathrm{m}$=2.37), which is still consistent with but slightly steeper than the Rayleigh scattering slope. 

\subsection{Transmission spectroscopy as a whole}

The transmission spectra of \object{GJ 3470b} have also been obtained by {\it HST}/WFC3 in the 1.1--1.7~$\mu$m band \citep{2014A&A...570A..89E} and by Keck/MOSFIRE in the 2.09--2.36~$\mu$m band \citep{2013A&A...559A..33C}. These two near-infrared transmission spectra have the same overall transit-depth level, and both indicate a flat featureless spectrum originating from high-altitude clouds or high metallicity atmospheric compositions. \citet{2014A&A...570A..89E} discussed in depth the prompted dichotomic spectrum when collecting the optical photometric measurements together with the near-infrared transmission spectrum. They found that the combination of the tentative Rayleigh scattering slope in the optical and the flat featureless spectrum in the near-infrared had to be explained by a cloudy hydrogen-rich atmosphere with an extremely low water volume mixing ratio, which, however, were not necessarily physically or chemically realistic \citep[e.g.][]{2014ApJ...784...63H}. 

\begin{figure}
 \centering
 \includegraphics[width=\linewidth,angle=0.0]{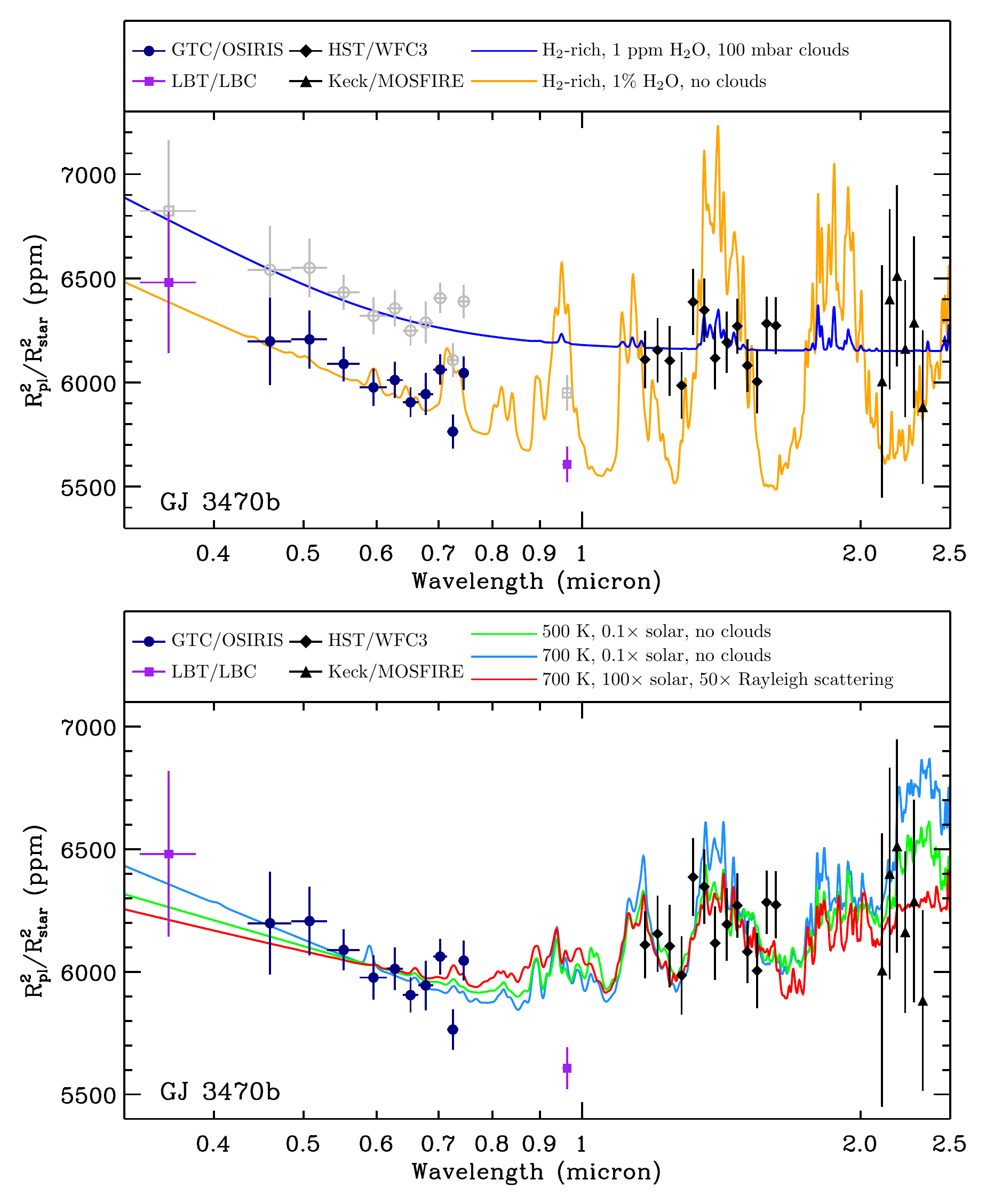}
 \caption{Transmission spectrum of \object{GJ 3470b}, which includes the transit depth measurements from GTC/OSIRIS (navy-blue circles), LBT/LBC (purple squares), {\it HST}/WFC3 (black diamonds), and Keck/MOSFIRE (black triangles). 
The two models shown in the {\it top panel} are taken from \citet{2014A&A...570A..89E}, which correspond to H$_2$-rich atmospheres with 1~ppm H$_2$O and clouds at 100~mbar (blue) or with 1\% H$_2$O and no clouds (orange), respectively. The three models shown in the {\it bottom panel} are computed using the \texttt{Exo-Transmit} code \citep{2016arXiv161103871K}. In the {\it top panel}, the GTC/OSIRIS+LBT/LBC measurements are shifted upwards 343~ppm, and then shown as empty symbols in gray.\label{fig:transpec}}
\end{figure}

When comparing to the {\it HST}/WFC3 near-infrared transmission spectrum, the GTC/OSIRIS and LBT/LBC combined optical transmission spectrum is clearly slightly downwards offset (see the top panel of Fig.~\ref{fig:transpec}). Planets with larger infrared radii than the optical would indicate predominantly clear atmospheres \citep[e.g.][]{2016Natur.529...59S}. However, in the case of \object{GJ 3470b}, the infrared measurements might suggest a cloudy atmosphere or a pure water atmosphere. While the Rayleigh scattering slope is spectroscopically confirmed, it is difficult to explain this optical trend that goes into atmospheric layers deeper than the cloud deck or the pure water absorption suggested by the infrared data. We note that the GTC/OSIRIS + LBT/LBC spectrum, except for the one in the $F972N20$ filter, well matches the prediction of a cloudy hydrogen-rich atmosphere \citep[i.e. the blue model in the top panel of Fig.~\ref{fig:transpec}; taken from][]{2014A&A...570A..89E} if they had been moved upwards $\Delta R^2_{\rm p}/R^2_\star=343$~ppm. 

While stellar activity remains a possibility to introduce an offset between data sets obtained at different epochs, we also note that our GTC/OSIRIS transmission spectrum is compatible with the HST/WFC3 measurements. In the bottom panel of Fig.~\ref{fig:transpec}, we show three atmospheric models computed using the \texttt{Exo-Transmit} code \citep{2016arXiv161103871K}: (1) a 700~K cloud-free model with 0.1$\times$ solar composition; (2) a 500~K cloud-free model with 0.1$\times$ solar composition; (3) a 700~K model with 100$\times$ solar composition and 50$\times$ enhanced Rayleigh scattering. Comparing these three models to the GTC/OSIRIS and HST/WFC3 combined data sets, the reduced chi-squares $\chi^2_{\rm r}$ are 1.60, 1.19, 1.17, respectively. In contrast, a horizontal line would result in $\chi^2_{\rm r}=1.78$. This indicates that the combination of lower temperature and lower metallicity in a cloud-free atmosphere can reduce the near-infrared spectral modulation while keeping a slope similar to the one suggested by the optical data. Similarly, it can also be achieved by higher metallicity with enhanced Rayleigh scattering from hazes. With the current data collection, the compositions of \object{GJ 3470b}'s atmosphere are still degenerate. We expect that the atmosphere of \object{GJ 3470b} can be well constrained in the future with more transit spectroscopy observations at higher precision, hopefully simultaneously in the optical and infrared.

\section{Conclusions}
\label{sectVI}

We have observed three transits of the warm sub-Uranus-mass planet \object{GJ 3470b} with GTC/OSIRIS in low-resolution spectroscopy. While the first transit was performed in extremely poor weather condition, the remaining two transits allowed us to obtain the first high-quality transmission spectrum in the wavelength range 435--755~nm. We were able to spectrally-resolve the Rayleigh scattering slope in \object{GJ 3470b}'s atmosphere for the first time. 

We have also observed one transit of \object{GJ 3470b} with VLT/UVES in high-resolution spectroscopy. We found that GJ 3470 is an active star with a magnetic filling factor of around 10--15\%. The relative flux of the emission cores of the \ion{Ca}{II}\,H+K lines changes by only $0.67\pm0.22$\% in- and out-of- transit. The relative flux of the H$\alpha$ lines changes by only $0.21\pm0.19\%$. We have not statistically detected any narrow absorption signatures originating from the planetary atmosphere.

Combining the low- and high-resolution spectroscopy, we have discussed the possible impacts from stellar activity. With current knowledge of the host star, the transmission spectrum is not likely significantly contaminated by the wavelength dependent effect from stellar activity. However, an overall shift from epoch to epoch caused by stellar activity is still possible. Therefore, a better observing strategy for future transmission spectroscopy of this planet would be a large wavelength coverage encompassed by a long-term photometric monitoring campaign.

\begin{acknowledgements}
    We thank the anonymous referee for his/her careful reading and helpful comments that improved the manuscript.
    We thank Dr. D. Ehrenreich for kindly providing us the theoretical atmospheric models. 
    This research is based on observations made with the Gran Telescopio Canarias (GTC), installed in the Spanish Observatorio del Roque de los Muchachos, operated on the island of La Palma by the Instituto de Astrof\'isica de Canarias.  
    This work is partly financed by the Spanish Ministry of Economics and Competitiveness through projects ESP2013-48391-C4-2-R, ESP2014-57495-C2-1-R, and AYA2012-39346-C02-02. 
    This work was supported by the Th\"uringer Ministerium f\"ur Wirtschaft, Wissenschaft und Digitale Gesellschaft. 
    G.C. also acknowledges the support by the National Natural Science Foundation of China (Grant No. 11503088) and the Natural Science Foundation of Jiangsu Province (Grant No. BK20151051).
    This research has made use of the VizieR catalogue access tool, CDS, Strasbourg, France \citep{2000A&AS..143....9W}.
\end{acknowledgements}

\bibliographystyle{aa} 
\bibliography{ref_db} 

\clearpage
\end{document}